\documentclass[%
 aip,
 jmp,%
 amsmath,amssymb,
reprint,%
]{revtex4-1}

\usepackage{graphicx}
\usepackage{dcolumn}
\usepackage{bm}
\usepackage[usenames]{color}
\definecolor{cRef1}{rgb}{1.0,0.0,0.0}

\begin{document}


\title{Revision of the Coulomb logarithm in the ideal plasma}

\author{P. Mulser}\email{peter.mulser@physik.tu-darmstadt.de.}
\affiliation{Institut f\"ur Angewandte Physik,
Technische Universit\"{a}t Darmstadt, D-64289 Darmstadt, Germany}
\author{G. Alber}
\affiliation{Institut f\"ur Angewandte Physik,
Technische Universit\"{a}t Darmstadt, D-64289 Darmstadt, Germany
}%
\author{M. Murakami}
\affiliation{ ILE: Institute of Laser Engineering, Osaka
University, Yamada kami, Osaka, Japan}

\date{\today}

\begin{abstract}
The standard picture of the Coulomb logarithm in the ideal plasma is controversial, the arguments for the lower cut off  need revision. The two cases of far subthermal and of far superthermal electron drift  motions are accessible to a rigorous analytical treatment. We show that  the lower cut off $b_{\min}$ is a function of symmetry and shape of the shielding cloud, it is not universal. In the subthermal case  shielding is spherical and $b_{\min}$ is to be identified with the de Broglie wavelength; at superthermal drift the shielding cloud exhibits cylindrical (axial) symmetry and $b_{\min}$ is the classical parameter of perpendicular deflection. In both situations the cut offs are determined by the electron-ion encounters at large collision parameters. This is in net contrast to the governing standard interpretation that attributes  $b_{\min}$ to the Coulomb singularity at vanishing collision parameters $b$ and, consequently, assigns it universal validity. The origin of the contradictions in the traditional picture is analyzed.
\end{abstract}


\maketitle

\section{Introduction}

In transport theory of the ideal plasma the Coulomb logarithm $\ln \Lambda$ plays a key role. Ohmic heating of a plasma, e.g. in the Tokamak, heating by microwaves or, at higher frequencies, collisional absorption of the high power laser beam, are all based on the same principle of directed momentum loss of the electron fluid to the ions by Coulomb collisions and conversion of kinetic electron fluid energy into electron heat. In this scattering process the single electron undergoes an elastic free-free transition accompanied by a deflection angle $\vartheta$. The electron-ion encounters are characterized by the three parameters of the screening or Debye length $\lambda_D$, the impact or collision parameter for perpendicular deflection $b_{\bot}$, and the reduced de Broglie wavelength $\lambda_B$. A plasma is considered non degenerate ideal if the two ratios $\lambda_D/b_{\bot}$ and $\lambda_D/\lambda_B$ are both much larger than unity. As a consequence the overwhelming number of encounters suffers small angle defections $\vartheta$ only with the implications that transport can be described in a linearized version and the superposition principle for simultaneous events holds; it allows to reduce simultaneous collisions to a sequence of binary interactions. In terms of trajectories this is equivalent to the two assumptions (i) of the existence of classical orbits and that (ii) in zero approximation the orbits are straight lines. Under extension of these hypotheses (small angle deflections or straight orbits) to {\textit{all} Coulomb collisions the calculation of the mean momentum transfer unavoidably ends in the so-called Coulomb logarithm, either in $b$ or $\vartheta$,
\begin{equation}
\ln\Lambda = \int \frac{db}{b} = \int \frac{d\sin\vartheta}{\sin\vartheta}. \label{ln}
\end{equation}
Both integrals diverge for $b \rightarrow 0$, corresponding to $\vartheta \rightarrow \pi$, and for $b \rightarrow \infty$, corresponding to $\vartheta \rightarrow 0$. The divergences have to be removed by the introduction of appropriate upper and lower "cut offs" $b_{\max}$ and $b_{\min}$. In justifying them rules have been introduced on the basis of "physical" arguments seven decades ago (or earlier) and subsequently they have become standard and have since then appeared in scientific papers time and again. A critical revision is in order.
\subsection{The Coulomb paradox}
The upper of the two cut offs  $b_{\min}, b_{\max}$ is well understood and not controversial.  When an electron of impact parameter $b$ collides, say with an ion, and the 'spectator' electrons with impact parameters $b' < b$ interact also with this ion at the same time the bare Coulomb potential is weakened and reduces to finite extent $r = b_{\max}$ of the range of the Debye length $\lambda_D$. Thus, the Coulomb potential has to be "cut off" at $b_{\max} = \lambda_D$. The number of simultaneous events is of the order of $\Lambda$. In magnetic fusion plasmas $\ln\Lambda$ ranges typically between $10$ and $20$ corresponding to $\Lambda \sim 2 \times 10^4 - 5 \times 10^8$. Characteristic values in laser plasmas are $\ln\Lambda \simeq 3 - 7$, i.e., $\Lambda \simeq 20 - 10^3$.\\

All difficulty concentrates on the lower cut off $b_{\min}$. A survey of the pertinent literature seems to support the existence of two groups. Statistically, $60\%$ of researchers apparently adhere to the setting  $b_{\rm min} = \lambda_B$. The standard "physical" motivation for this lower cut off is that an orbit cannot be localized better than the de Broglie distance and therefore impact parameters $b < \lambda_B$ are meaningless. However, in the case $b_{\bot}$ exceeds $\lambda_B$, $b_{\min}$ should be identified with the impact parameter $b$ of "closest approach" $b_{\bot}$ because in this situation such a classical orbit has a well defined meaning. The majority of representative textbooks and specialized papers, e.g. \cite{wiki}, adhere to this hypothesis which can be summarized quantitatively as
\begin{equation}
b_{\rm min} = \max\{\lambda_B, b_{\bot}\}; \hspace{0.2cm}
\lambda_B = \frac{\hbar}{m_{\rm e}v_r},\,\, b_{\bot} = \frac{Ze^2}{8\pi\varepsilon_{\rm 0}E_r}    \label{bmin}
\end{equation}
with $m_{\rm e}$ electron mass, $v_r$, $E_r$ relative velocity and energy between encounters, $e$, $Z$ elementary charge and ion charge number, $\varepsilon_0$ dielectric constant. Note, $\lambda_B \sim E_r^{-1/2}$, $b_{\bot} \sim E_r^{-1}$. Numerically $\lambda_B, b_{\bot}$ result as
\begin{equation}
\lambda_B [{\rm nm}] = \frac{0.185}{(E_r [{\rm eV}])^{1/2}},\hspace{0.5cm} b_{\bot} [{\rm nm}] = \frac{0.7\times Z}{E_r [{\rm eV}]}. \label{num}
\end{equation}
A special argument for setting eq.(\ref{bmin}) is by L. Spitzer \cite{spitz}. He arrives at the limitation $b \geq \lambda_B$ by observing that for impact parameters $b \leq \lambda_B$ the Coulomb differential cross section leads  to higher diffraction values than an opaque disc of the same radius, which is "unphysical". It seems that for numerous researchers this constitutes the basic argument.
The second, minor group, to reference a few examples \cite{rosen} $^-$ \cite{khrap}, believes that $b_{\min} = b_{\bot}$ is the correct setting and collects various arguments for it, for example the argument of the closest approach.\\
\indent From the interpretation of the lower cut off of both groups a first consequence must be drawn: $b_{\min}$, either set equal to $\lambda_B$ or $b_{\bot}$, is universal in the sense that at the lower cut off the electron-ion interaction is Coulomb (or Rutherford) like and does not depend on other parameters, as for example the geometry of the screening cloud. Further, it is evident that at most only one of the settings can be correct. Least, the argument of closest approach is invalid because the minimum impact parameter in the center of mass system is zero. The introduction of cut offs can only be regarded as a recipe. If there exists a lower cut off it must be the result of the correct treatment of scattering to begin at $\vartheta = \pi$, resp. $b = 0$, and without making any use of the straight orbit approximation. By the latter linearization is excluded.\\
\indent One more consideration is in order; it leads to the proper Coulomb paradox in the restricted sense. Setting $b_{\min} = b_{\bot}$ results correctly from calculating the momentum transfer in a collision if the integration is done for a bare Coulomb potential from $b= 0$ up to its range $b = b_{\max} = \lambda_D$, \cite{rosen}. This is a genuine paradox if one keeps in mind that in the case of the ideal plasma the overwhelming number of orbits is classical and straight in zero approximation, and that the bent orbits of close encounters follow the correct Rutherford differential scattering cross section. Here one could, and one must argue that the bare Coulomb potential has to be replaced by the Debye potential because at $b = \lambda_D$ the two potentials differ substantially from each other. The outcome of the calculation reproduces the result from before, i.e.,  $\ln\Lambda = \ln(\lambda_D/b_{\bot})$. Hence, we are dealing with a genuine paradox. In conclusion, the situation is controversial, the standard interpretation of $b_{\min}$ as a "lower cut off" and the arguments used for specific values of it are inconsistent and self-contradictory. Clarification on a solid basis is needed.

\subsection{The role of a quantum treatment}

The setting $b_{\min} = \lambda_D$, if true, must have its roots in quantum mechanics; $b_{\min} = b_{\bot}$ is based on classical mechanics of well defined orbits. It is evident, and it is generally accepted, that in case of divergent results the quantum treatment applies and the classical result must be rejected. In the following determination of the correct Coulomb logarithm we strictly adhere to quantum mechanics although there may apparently exist good reasons for a classical treatment, as believed for example by the authors of ref. \cite{rosen} and many others.\\
\indent There is no general rule for a legitimate transition to the simpler classical model. In one and the same problem it depends on the magnitude or variable under consideration whether the conditions for a classical analysis are fulfilled. Examples are the equation of state of an ideal gas and its mixing entropy; the first is classical, the second quantity follows the quantum Sackur-Tetrode formula and not its classical counterpart; or the electron partition function in the Saha equation which at arbitrarily high temperature and infinite dilution does not assume the classical value. A particularly simple example is the pressure exerted by a beam of independent particles of momentum ${\bf p} =$ const. In both cases it is $(2)n|{\bf p}|$, $n$ particle density. A further example is the connection between pressure $p$ and energy density $\epsilon$ of an ideal gas of $f = 3$ degrees of freedom in thermal equilibrium,  $p/\epsilon = f/(f + 2)$ for non-relativistic cold Fermi and classical gases, and $p/\epsilon = f/(f + 1)$ for them and for photons in the superrelativistic regime. By the way, a photon number state $|n>$ never turns over into "classical light" regardless of how large the photon number is in the mode \cite{loud}.\\

In this paper we present the binary scattering problem for point charges by strictly adhering to the quantum view in two relevant cases, one for vanishing subthermal electron drift $u << v_{\rm th}$, $v_{\rm th}$ electron thermal velocity, the other one for far superthermal flow $u >> v_{\rm th}$. Starting from the two effective Hamiltonians we shall get the correct "lower cut offs" by subjecting them to the standard quantum procedure and to arrive at a coherent interpretation of their meaning that is free of contradictions. On the way to the solution it will appear essential to distinguish between the validity of classical mechanics for the \textit{single} orbit as the limiting quantum case and the correctness of this passage when properties of orbits are to be determined that are the result of folding on the \textit{totality} of the orbits. It allows for a complete revision of the subject.The result will surprise: There is no cut off; $b_{\min}$ is the result of the charge distributions at large impact parameters and will depend on the symmetry/geometry of the screening cloud.
\section{Shielding symmetry determines lower cut off}
\subsection{Spherical shielding}

A universally accepted fact by the entire plasma community is that in a plasma with isotropic monoenergetic electron distribution function $f({\bf w}) = v\delta(|{\bf w}| - v)/V$, $V$ volume, the effective potential of an ion in random phase approximation (RPA) is of the Debye/Yukawa potential type $\Phi_s$ with screening length $\lambda_s$
\begin{eqnarray}
\Phi_s = \frac{q}{4\pi \varepsilon_0 r} \exp{(-r/\lambda_s)};\hspace{0.4cm}
\lambda_{s} = \big(\frac{\varepsilon_0 \mu}{n_{\rm e}
e^2 1/v^2}\big)^{1/2}        \label{screen}
\end{eqnarray}
reduced mass $\mu \simeq m_{\rm e}$, electron density $n_{\rm e}$. For $f({\bf v})$ Maxwellian, $<1/\mu v^2> = 1/k_B T_e$, screening results into the thermal Debye potential $\Phi_D$ with range of the Debye length $\lambda_D$,
\begin{eqnarray}
\Phi_D = \frac{q}{4\pi \varepsilon_0 r} \exp{(-r/\lambda_D)};
~~\lambda_D  = \big(\frac{\varepsilon_0 k_BT_{\rm e}}{n_{\rm e}e^2}\big)^{1/2}
\label{debye}
\end{eqnarray}
$k_B$ Boltzmann constant, $T_{\rm e}$ electron temperature. Classical and quantum mechanical derivation, e.g. along \cite{roep}, lead to identical results for (\ref{screen}) and (\ref{debye}). In the derivation of $\Phi_s$ solely the law for the electrostatic force ${\bf f} = q{\bf E}$ on the point charge $q$ by the field ${\bf E}$ and Poisson's equation are needed, as for the derivation of the bare Coulomb potential $\Phi_C$. To keep the analysis as lucid and simple as possible without loss of generality it is indicated to limit to the isotropic monoenergetic electron distribution and to concentrate on the case study of the average momentum transfer $\Delta{\bf p}$ from the electron of velocity ${\bf v}$ to the ion,
\begin{equation}
\Delta{\bf p} = 2\pi \mu{\bf v} \int_0^\infty \frac{\sigma_s(b)}{\sigma_0}(1 - \cos\vartheta)\,db;\,\, \sigma_0 =  \int \sigma_s(b)2\pi b \,db. \label{momentum}
\end{equation}
The effective Hamiltonian is
\begin{equation}
H({\bf p}, {\bf r}) = \frac{{\bf p}^2}{2\mu} + \Phi_s(r), \hspace{0.5cm}r = |{\bf r}|. \label{ham1}
\end{equation}
For a Debye-like potential standard scattering theory then yields the differential cross section $\sigma_s$ in first Born approximation for free-free transitions \cite{sakur},
\begin{equation}
\sigma_s(\vartheta) = \frac{b_{\bot}^2}{4[\sin ^2(\vartheta/2) +
(\lambda_B/2\lambda_s)^2]^2}; \hspace{0.4cm} \tan\frac{\vartheta}{2} = \frac{b_{\bot}}{b}.             \label{born}
\end{equation}
The term $\rho = \lambda_B/2\lambda_s$ in $\sigma_s$
is the contribution from the shielding factor
$\exp(-r/\lambda_s)$ to the Fourier transform of the Debye potential. For $\lambda_s = \infty$, $\sigma_s$ shrinks to the well-known Rutherford or bare Coulomb potential $\sigma_C$.  For large $k \Leftrightarrow {\rm small}\, b \sim 1/k$ only the $\{r/\lambda_s << 1\} $ region contributes and hence $\sigma_s = \sigma_C $. For small $k$ the outer region $ r \simeq \lambda_s $
counts where $ \cos \vartheta = 1 - \vartheta^2 /2 \lesssim  1 -
(\lambda_B / 2 \lambda_s)^2 $. Eq. (\ref{born}) inserted in (\ref{momentum}) and integrated from $b = 0 \Leftrightarrow \vartheta = \pi$ to $b = \infty \Leftrightarrow \vartheta = 0$ the Coulomb logarithm $L_C$ results,
\begin{eqnarray}
L_C = \frac{1}{2}\big[\ln\big(\frac{1}{\rho^2} + 1 \big) - \frac{1}{1 +
\rho^2}\big] \simeq \frac{1}{2}\big(\ln\frac{1}{\rho^2} + 2\rho^2 - 1)    \nonumber
\end{eqnarray}
\begin{eqnarray}
= \ln \frac{\lambda_s}{\lambda_B} + 0.2 +
\frac{\lambda_B^2}{4\lambda_s^2}. \label{LC}
\end{eqnarray}
Note, $b$ is an integration variable only, related to the scattering angle $\vartheta$ by (\ref{born}), and not affected by whether there exist classical orbits or not. To obtain the usual average momentum transfer in thermal equilibrium folding over the Maxwellian has to be done (as for instance in \cite{schneider}). Then, if  the logarithmic expression does not depend much on the velocity, $\lambda_s$ and $\lambda_B$ are taken out of the integral and approximated by the thermal Debye and the thermal de Broglie lengths. However, this last step must be checked from case to case (for instance in laser plasmas with supergaussian velocity distributions \cite{lang} or in planetary plasmas with a $\kappa$ distribution \cite{kappa}). It has no influence on our analysis with the monoenergetic distribution.\\
\indent By (\ref{LC}) we have shown that in the ideal non degenerate plasma with $\lambda_D >> \lambda_B$ the correct lower cutoff is the de Broglie length and not the parameter for perpendicular deflection, in contradiction to \cite{rosen}, \cite{daw} and others. The cut offs $b_{\min}$ and $b_{\max}$ are the result of integration of the impact parameter from 0 to $\infty$; no additional physical hypotheses are needed. The integral is regular in the whole domain, the neighborhood of $r = 0$ is in no way special. Finally, there is no basis for such a rule as $b_{\rm min} = \max\{\lambda_B, b_{\bot}\}$. It is a mere guess, no proof has ever been given. The situation with $b_{\bot} >> \lambda_B$ is more complicated and not analyzed here. The analysis given here leads to a completely different interpretation of why the de Broglie length comes into play in the Coulomb logarithm. Inspection of (\ref{born}) shows that for small $k$'s, i.e., large $b$'s the outer region $r \simeq \lambda_s$ counts where $\cos\vartheta = 1 - \vartheta^2/2 \lesssim 1 -(\lambda_B/2\lambda_s)^2$. Hence, contrary to the dominating interpretation of $\ln \Lambda$ screening due to the outer regions is responsible for the "lower cut off" and not the singularity of the Coulomb potential at $r = 0$.\\
\indent Thanks to a hint by S. Atzeni \cite{atzeni} we have realized very recently that there exists one treatment of the Coulomb logarithm in which the author comes to the conclusion identical to ours that $b_{\min} = \lambda_B$ originates from shielding at large impact parameters $b$ [but for the rest he still adheres to the traditional picture, e.g., setting (\ref{bmin}) for $b_{\min}$]. It is found in the representative textbook \textit{Tokamaks} by John Wesson \cite{wesson} and up to now it is the only statement among all papers known to us. Apparently its impact has been almost zero so far.
\subsubsection{Validity of the first Born approximation}
Let us start with two qualitative arguments. For the first Born approximation to be correct the local, partially scattered state function $\psi({\bf r})$ should be close to the incident wave $\phi({\bf r})$ everywhere. In other words, scattering must be weak for all angles $\vartheta$. We can assume that with the Debye potential this is true for the following reason. The Debye potential $\Phi_s$ is smooth and weaker than the bare Coulomb potential in  the whole region. The Coulomb-Rutherford cross section $\sigma_C$ is correct to all orders and it agrees with its first Born approximation (see for instance \cite{sakur}). Thus, the condition for its use in $\sigma_s$ is mathematically fulfilled if this argument extends onto real and imaginary part contributing to the modulus of the scattering amplitude separately. Owing to the smooth transition between the two scattering potentials Coulomb $\Leftrightarrow$ Debye this is very likely to be the case. The perfect analogy to classical optics for diffraction from spatial filters may help to convince.\\
\indent Another qualitative argument is obtained from considering the attenuation of a plane  wave of momentum ${\bf p}$ incident onto a homogeneously distributed ensemble of ions of density $n$. The attenuation follows Beer's exponential law $I = I_0\exp[-<n\sigma>x]$ with $<n\sigma> = 4\pi b_{\bot}^2n\ln\Lambda$ from (\ref{momentum}), (\ref{born}) and (\ref{LC}). The attenuation by a  monoionic layer of thickness $\Delta x = n^{-1/3}$ and the representative numerical example $n = 10^{21}{\rm cm}^{-3}, E_r = 100$ eV tells that the first Born approximation is very well fulfilled,
\begin{equation}
I = I_0\exp[-<n\sigma>\Delta x];\hspace{0.5cm}<n\sigma> \Delta x = 4\pi b_{\bot}^2n^{2/3}\ln\Lambda  \nonumber
\end{equation}
\begin{equation}
= 3Z^2\times 10^{-5}; \hspace{0.5cm}\ln\Lambda = 4.9. \label{qualcrit}
\end{equation}
The qualitative arguments show tendencies in the parameter region. \\
\indent A rigorous criterion is obtained from wave packet considerations with natural transverse dispersion
\begin{equation}
\phi({\bf r}, t) = (\pi L_t^2)^{-3/4}\exp-\frac{({\bf r} - {\bf v}t)^2}{2L_t^2}; \hspace{0.3cm} L_t^2 = L^2 +\frac{\hbar^2t^2}{\mu^2L^2}.  \nonumber
\end{equation}
Following the mathematical analysis by \cite{farina} for the Debye potential $\phi_s$ we determine the ratio of the two moduli
\begin{equation}
\phi_1 = \lambda_D^{-3}|\int\phi_s[r]dr| = 4\pi \frac{\kappa}{\lambda_D}, \phi_2 = \big(\lambda_D^{-3}\int|\phi_s(r)|^2dr\big)^{1/2} \nonumber
\end{equation}
\begin{equation}
= (8\pi)^{-1/2} \phi_1; \hspace{0.5cm} \kappa = \frac{Ze^2}{4\pi \varepsilon_0}.  \nonumber
\end{equation}
Let $\alpha = |\psi(0) - \phi(0)|/|\phi(0)|$ be the relative error. Then, with $L = (\lambda_B \lambda_D)^{1/2}$ it is bound by
\begin{equation}
\alpha \leq \big[\frac{\pi^{1/4}}{\sqrt{2}}\big(\frac {\lambda_B}{\lambda_D}\big)^{1/4} +1\big]\frac{\phi_2}{\phi_1}\frac{b_{\bot}}{\lambda_B} = 0.2\big[0.9\big(\frac {\lambda_B}{\lambda_D}\big)^{1/4} +1\big]\frac{b_{\bot}}{\lambda_B}; \nonumber
\end{equation}
\begin{equation}
\alpha \hspace{0.3cm} \rightarrow \hspace{0.3cm}0.2\frac{b_{\bot}}{\lambda_B}. \label{borncrit}
\end{equation}
For our case from above, i.e., $n = 10^{21}$cm$^{-3}$, $E_r = 100$ eV, results
\begin{equation}
\frac{L}{\lambda_B} =  11.3, \hspace{0.2cm}  \big(\frac{\lambda_B}{\lambda_D}\big)^{1/4} = 0.3, \hspace{0.2cm} \frac{b_{\bot}}{\lambda_B} = 0.38; \hspace{0.2cm} \alpha = 0.1. \nonumber
\end{equation}
Thus, for this situation close to ideality the use of the first Born approximation is legitimate. For the Tokamak plasma $\alpha $ is of the order of $10^{-2}$. Instead of following the demanding wave packet analysis of \cite{farina} one could think of proceeding to the second and higher Born approximations. It is not feasible because the second Born approximation already diverges.

\subsection{Cylindrical screening}
The second case accessible to an analytical treatment and of high relevance in applications is that of far superthermal drift velocity $|{\bf u}| >> v_{\rm th}$. The electron distribution function is assumed as $f({\bf w}) = \delta({\bf w - v})/V$. In such an electron fluid flow the interaction with an ion can be seen as the ion moving with velocity $-{\bf v}$ through the cold electron fluid at rest. The disturbance caused by the interaction appears as a polarization wake, or in other words, as Cherenkov emission of plasmons. Let us characterize the polarization as ${\bf P} = - n_{\rm e}e{\boldsymbol \delta }$, with ${\boldsymbol \delta}(b)$ the displacement of the electrons from their equilibrium position. By applying exactly identical physics, particle conservation and Poisson's equation ,
\begin{equation}
\nabla n_e + n_e\nabla{\boldsymbol \delta } = \nabla\big(- \frac{\varepsilon_0}{e}{\bf E} +n_e {\boldsymbol \delta } \big) = Ze\delta({\bf r}), \nonumber
\end{equation}
plus force law ${\bf f} = q{\bf E} = \mu \ddot{{\boldsymbol \delta }}$ as in the former case of spherical far subsonic screening one arrives, quantum  mechanically as well as classically, at Bohr's celebrated oscillator model for ${\boldsymbol \delta} \sim {\bf P}$, \cite{sigmund}
\begin{equation}
{\bf\ddot{\bm \delta}} + \omega_p^2{\bm \delta} = {\bf f}_C/\mu; \hspace{0.4cm} {\bf f}_C  = \frac{Ze^2}{4\pi\varepsilon_0r^2},
\label{oscillator}
\end{equation}
with the plasma frequency $\omega_p = (n_{\rm e}e^2/\varepsilon_0\mu)^{1/2}$. The ion of charge $q = Ze$ is supposed to sit at ${\bf r} = 0$. The oscillator term $\omega_p^2{\bm \delta}$ provides for dynamic shielding. An electron starting at $x = - \infty$ is attracted by the ion as it comes closer thereby reducing its collision parameter $b$ to $b' < b$. As a consequence,  the electron density increases from $n_e$ to $n_e' =n_eb/b'$ and creates the restoring force in (\ref{oscillator}). At $x = + \infty$ the electron is free again and is left in an excited oscillation state. The oscillation energy occurs on the expense of kinetic energy of the ion. Bohr used the model to calculate ion beam stopping in ionized matter. In general it applies to strong drift motions under negligible transverse temperature, e.g., fast electron transport in laser plasmas \cite{macchi}.\\
\indent Adherent to our principle we subject the equation of motion (\ref{oscillator}) to a quantum treatment by looking for the corresponding Hamiltonian. It reads, with the Coulomb interaction in dipole approximation $(eq/4\pi \varepsilon_0){\bm \delta}\nabla(1/r) = {\bf f}_C(vt, b)\,{\bm \delta}$,
\begin{eqnarray}
H({\bf p}, {\bm \delta}_{\rm op}, t) = \frac{{\bf p}^2}{2\mu} + \frac{\mu}{2}\omega_p^2\,{\bm \delta}_{\rm op}^2 - {\bf f}_C\,{\bm \delta}_{\rm op} = H_0 + H_C. \hspace{0.4cm}\label{ham2}
\end{eqnarray}
Index "op" stands for operator. The solution is given in terms of coherent or Glauber states (see e.g., \cite{tan} or \cite{loud}). The ground state $|\psi_i> = |0>$ at $t = - \infty$ is driven by $H_C$ into the coherent Glauber eigenstate $|\psi_f> = {|\hat{\bm \delta}>}$ at $t = +\infty$. For obvious reasons it is labeled here by the classical amplitude ${\hat{\bm \delta}}$: the expectation value $<\psi_f|{\hat{\bm \delta}}_{\rm op}|\psi_f>$ of the asymptotic shift at $t = + \infty$ coincides with its classical value ${\hat{\bm \delta}}$ from eq.(\ref{oscillator}). The solution of ${\bm \delta} = (\delta_{\bot}, \delta_{\|})$ is
given in terms of the modified Bessel functions $K_1$ and $K_0$
\cite{abram}, with the amplitudes  \cite{arnold}
\begin{eqnarray}
\hat{\delta}_{\bot}(\beta) = 2b_{\bot}K_1(\beta),
\hat{\delta}_{\|}(\beta) = 2b_{\bot}K_0(\beta);  \beta = b/\lambda,
\lambda = v/\omega_p. \nonumber
\end{eqnarray}
For $b$ small,  $K_1$ and $K_0$ diverge both as a
consequence of the linearization in polarization ${\bf P}$. For vanishing impact parameters $b$, $\omega_p$ reduces smoothly to zero owing to missing screening and interaction goes over into bare Coulomb scattering, as in the former subthermal case with $\Phi_s$. Therefore regularization is done  by integrating the oscillator solution from $ b_0 = sb_{\bot} << \lambda$ to infinity, factor $s > 5$, and treating the momentum transfer $D(\beta_0)$ of the close encounters in $0 \leq b \leq b_0$ by scattering from the unscreened Coulomb potential or, with the same result, from (\ref{born}). Then the total energy $\dot{W}$ irradiated into plasmons per unit time is
\begin{eqnarray}
\dot{W} = \frac{1}{2}\mu\omega_p^2v\lambda^2\int_{\beta_0}^\infty 2\pi \beta(\hat{\delta}_{\bot}^2 + \hat{\delta}_{\|}^2)d\beta +
v^2D(\beta_0)\nonumber
\\=4\pi\mu v^3b_{\bot}^2\big[\beta_0K_0(\beta_0)K_1(\beta_0) +
\frac{1}{2}\ln\frac{b_0^2 + b_{\bot}^2}{b_{\bot}^2} \big].
\label{work}
\end{eqnarray}
Thereby use has been made of $d(\beta K_0 K_1)/d\beta = -
\beta(K_0^2+ K_1^2)$. The integrals $\int\beta K_0^2d\beta$, $\int\beta
K_1^2d\beta$ and their sum are shown as functions of $\beta$ in Fig. 1.
\begin{figure}[h]
  \centering
  \vspace{0.3cm}
  \includegraphics[angle=0,width=0.8\columnwidth]{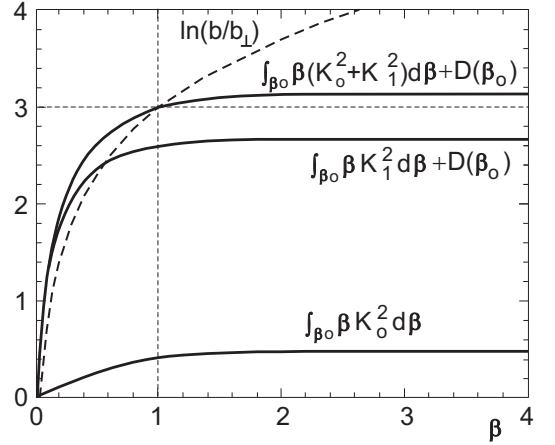}
  \caption{Oscillator model:
Transverse and longitudinal oscillation energies $E_{\parallel}\sim\int\beta
K_0^2d\beta $, $E_{\perp}\sim \int\beta K_1^2d\beta$ and their sum as functions of $\beta = b/\lambda$, ~$\lambda =
v/\omega_p$. The Coulomb logarithm $\ln\Lambda =
\ln(\lambda/b_{\perp})$ is a good approximation to $E_{\perp}(\beta
= \infty)$ even at the value as low as $\ln\Lambda = 3$; the small
deviation derives from regularization $\ln(1 + b_{\bot}/b_0)$. For
$b_0$ see text. Both, $K_0$ and $K_1$ diverge for $\beta \rightarrow
0$ and so does $\int\beta K_1^2d\beta$; ~$\int\beta K_0^2d\beta $ is
finite. For $\beta$ large $\rightarrow \beta K_0K_1 \sim
\exp-2\beta $.}
  \label{Fig.1.eps}
\end{figure}
Although the associated Coulomb logarithm, determined below,  is as low as $\ln\Lambda = 3$ it is an excellent approximation to the saturation of  solution eq. (\ref{work}) which is due to the oscillatory term in the Hamiltonian (\ref{ham2}). The ratio of the longitudinal to the transverse asymptotic oscillation energy is
\begin{equation}
\frac{E_{\|}}{E_{\bot}} = \frac{1}{2\ln\Lambda}.     \label{ratio}
\end{equation}
\subsubsection{Regularization}
Again a qualitative consideration first may support clarity. The generalization  of the Hamiltonian eq.(\ref{ham2}) with continuous transition of $\omega_p \rightarrow 0$ is possible with the aid of the Jacobian $J = \partial({\boldsymbol \delta})/\partial({\bf r})$. The equation however, one ends up with is untractable. Help comes from selecting the cut $b = b_0$ in such a way that (i) the single orbits within $b > b_0$ become classical, i.e., the state vectors $|\psi(b)>$ are expressible in the form of an action integral, (ii) these orbits are sufficiently straight owing to ${\bf P} = -en_e{\boldsymbol \delta}$ linearized, $n_e =$ const, (iii) the oscillator term in $H_0$ is much smaller than the driver term $H_C$ in $b \leq b_0$, and (iiii) $b_0$ is placed in a region where its individual choice is insensitive within a wide range. Condition (ii) is fulfilled for $1 - \cos\vartheta \simeq \vartheta^2/2 <<1$, that is $s = b_0/b_{\bot} \geq 5 \Rightarrow \cos\vartheta \geq 0.92$. This is also the condition for the fulfillment of (i),see Sec. III or \cite{sakur}, p. 103. With the electron-ion interaction time $\tau = 2b/v$, \cite{jackson} one deduces (iii) that the oscillator term in the Hamiltonian becomes insignificant for the easy condition $b_0^2/\lambda^2 << 1/4$. In the ideal plasma requirements (i) - (iii) are simultaneously fulfilled  for the setting $b_0 = \sqrt{b_{\bot}\lambda}$ proposed by \cite{farina} for the width $L$ for wave packets. For example, $\ln\Lambda = 5$ yields $s = 12$; from $\ln\Lambda = 10$ follows $s = 150$. Finally, it is a fortuitous circumstance that for $b \rightarrow 0$ the driver $H_C$ prevails so strongly on the oscillator term in $H_0$ that $\omega_p =$ const or $\omega_p \rightarrow 0$ makes asymptotically no difference.\\
\indent From $\beta$ small follows $\beta K_0K_1
= - \ln\beta/2 \times [1 + (\beta^2/2)\ln\beta/2] - \gamma$, Euler constant $\gamma = 0.57722$, \cite{abram}. Applied to $b = b_0$ the generalized Coulomb logarithm $L_C$ from the square bracket in (\ref{work}) evolves into
\begin{equation}
L_C = \beta K_0(\beta_0)K_1(\beta_0) + D(\beta_0) \nonumber
\end{equation}
\begin{equation}
= \ln\frac{\lambda}{b_{\bot}} + \ln2 - \gamma - \ln s + \frac{1}{2}\ln\big[s^2\big(1 + \frac{1}{s^2}\big)\big] -
\frac{1}{2}\ln\frac{b_0^2 + b_{\bot}^2}{b_{\bot}^2} \nonumber
\end{equation}
\begin{equation}
- \frac{\beta_0^2}{2}\ln^2\frac{\beta_0^2}{2} = \ln\Lambda + 0.116 + \Delta.  \label{LC2}
\end{equation}
The two terms $- \ln b_0 \sim - \ln s $ and $\ln s^2/2$ cancel each other guaranteeing insensitivity with respect to the special choice of $b_0$. In fact, the difference $\Delta$ is
\begin{equation}
\Delta = \frac{1}{2}\ln\big(1 + \frac{1}{s^2}\big) + \frac{s^2}{2\Lambda^2}\big(\ln^2\Lambda - \ln^2 s\big).  \label{diff}
\end{equation}
At $\ln\Lambda = 5$ and $s = 12$ the correction amounts to $\Delta = 6.1\times 10^{-2}$. At $\ln\Lambda = 10$ and $s = 150$ results $\Delta = 1.7\times 10^{-3}$. For $s = 20$ and $200$ at $\ln\Lambda = 10$ the deviation is $\Delta = 3.7\times 10^{-5}$ and $3.0\times 10^{-3}$.\\

In summary we have shown that in the plasma with cylindrical (axial) symmetry of shielding the correct, i.e. quantum Coulomb logarithm is given in leading order by $\ln\Lambda = \ln(\lambda/b_{\bot})$ from (\ref{LC2}) with the lower cut off this time determined by the classical impact parameter of perpendicular deflection $b_{\min} = b_{\bot}$. The lower cut off is not universal, it depends on the geometry of the screening cloud.\\

\section{Discussion and conclusion}

For the subthermal case the authors \cite{rosen} - \cite{bal} come to the conclusion that $b_{\min} = b_{\bot}$ instead of $\lambda_B$: The overwhelming majority of orbits are straight and classical; the bent orbits close to the ion are scattered by the classical Rutherford cross section $\sigma_C$ which is identical to its quantum mechanical expression. Hence, $\sigma_C$ applies to the entire region and the classical outcome might appear stringent. On the other hand, the quantum treatment in first Born approximation has been shown very well fulfilled for the ideal non degenerate plasma. We are faced indeed with a case where classical and quantum analysis yield different results. The solution of this peculiar situation is as follows. In the analysis involving one or a finite number of orbits a classical approach may be sufficiently precise. However, when folding over \textit{all} orbits the single tiny deviation from the classical limit, always present due to the uncertainty relation, may accumulate to a sensitive error, as it evidently does: Whether a classical analysis applies depends on the character of the quantity to be determined. To the single orbit a WBK criterion (after Wenzel, Kramers, Brillouin) may apply; when used globally it needs a proof. \\
\indent There is Spitzer's diffraction argument \cite{spitz}, page 128. Although physically appealing at first glance it is false and self contradictory. In the neighborhood of the Coulomb singularity the author compares Rutherford scattering with optical diffraction from a diaphragm of  diameter $2\lambda_B$. In doing so he seems not to be aware of comparing electron scattering from the $1/r$ potential and from a potential $V(r) = - \infty$ for $r \leq \lambda_B$ and $V = 0$ outside. By defining the refractive index as $n^2({\bf x}) = [1 - V({\bf x})/E]$ the Schr\"odinger equation of energy $E$ becomes
\begin{equation}
\nabla^2\psi + {\bf k}_0^2 n^2 \psi = 0; \hspace{0.4cm} {\bf k}_0^2 = \frac{2\mu E}{\hbar^2}.    \label{schrod}
\end{equation}
Apart from polarization it is identical  with the wave equation of optics governing diffraction,
\begin{equation}
\nabla^2{\bf E} + {\bf k}_0^2 n^2({\bf x}){\bf E} = 0; \hspace{0.4cm}|{\bf k}_0| = \frac{\omega}{c}.
\end{equation}
Furthermore, the smooth Debye potential does not generate diffraction fringes; comparison with the diaphragm after all violates Babinet's principle. Finally, Schr\"odinger's equation is identical to the scalar Kirchhoff diffraction equation. Spitzer's setting  of  $b_{\min} = \lambda_B$ is a prominent example of excellent physical intuition but mistaken proof.\\
\indent The different outcome for $b_{\min}$ from the Debye potential and the harmonic oscillator equation is not so surprising at closer inspection. The Coulomb potential and the harmonic potential are the only ones for which an additional conservation quantity, the so-called Lenz-Runge vector exists. In classical dynamics it leads to closed orbits (known as Bertrand's theorem \cite{bert}), quantum dynamics  of them is characterized by higher symmetry, additional degeneracy, and absence of $\hbar$ in the differential scattering cross sections. In the Debye potential $\phi_s$ no closed orbits exist, angular momentum degeneracy disappears, and $\lambda_B \sim \hbar$ appears in the differential scattering cross section, eq.(\ref{born}).\\
\indent The analysis given here may help to extend the investigations further into a region of not so ideal plasmas and to develop analytical expressions to be used in numerical programs of collision codes. As here our aim has been to revise the concept of "cut offs" and to arrive at a contradiction-free interpretation of the Coulomb logarithm we want to point out to the reader that in principle our scope is perfectly reached if the most idealized cases are assumed: infinitesimal drift in the Debye potential and infinitesimal electron temperature in the oscillator model. For these cases of maximum ideality the small corrections to the Coulomb logarithm disappear asymptotically. \\
\indent We have treated binary electron-ion collisions. In the ideal plasma the standard situation is the simultaneous interaction of a huge number of collision partners. It is the role of kinetic theory to offer a systematic approach to reduce the simultaneous events to a succession of binary small angle encounters by introducing appropriate effective potentials providing for screening. Our work sets in here. Procedures like the BBGKY hierarchy (after Born, Bogoljubov, Green, Kirkwood, Yvon)\cite{radu} and the generalized Kadanoff-Baym technique\cite{kab1}, \cite{kab2}, \cite{sem} are well known efficient reduction methods to the point where we start. The very many analytical approaches to screening along these reduction schemes (classical dielectric procedure, Green's function technique, Lindhard's model, Hartree-Fock approximation, etc.) have the limitation to linearity and straight orbit approximation in common. It is this fact that leads to the divergence of $\ln \Lambda$ at vanishing impact parameters, followed by the necessity of introducing somehow a 'lower cut off' $b_{\min}$. Expansion into higher diagrams does not circumvent the divergence. Any singularity is avoided by allowing also bent orbits, as done in this paper.\\

In conclusion we have found that in the plasma not far from ideality  the 'lower cut off' $b_{\min}$ is not a universal property and not based on the uncertainty principle applied at $b = \lambda_B$. It has its origin in the scattering at large impact parameters; its value depends on the profile of the screening potential and on its geometry. Each screening potential exhibits its individual $b_{\min}$, spherical potential $b_{\min} = \lambda_B$,  axisymmetric screening $b_{\min} = b_{\bot}$. Reinterpretation of $b_{\min}$ leads to a coherent picture of the Coulomb logarithm in the ideal plasma. Our results may offer help in formulating more precise cut offs in numerical codes of collisional absorption.\\



\end{document}